\documentclass[12pt]{iopart}
\usepackage{epsfig}
\usepackage{graphics}
\usepackage{graphicx}
\usepackage{epstopdf}
\usepackage[dvipsnames]{xcolor}
\DeclareGraphicsExtensions{.jpeg,.eps}

\begin{document}

\title[Order of BEC in mean-field approximations]{On the order of BEC transition in weakly interacting gases predicted by mean-field theory}

\author{L. Olivares-Quiroz\footnote{Also at Universidad Aut\'onoma de la Ciudad de M\'exico, M\'exico, D.F., Mexico.} and V. Romero-Rochin\footnote{Corresponding author}}

\address{Instituto de F\'{\i}sica. Universidad Nacional Aut\'onoma de M\'exico. \\
Apartado Postal 20-364, 01000 M\'exico, D.F., Mexico}
\ead{olivares@fisica.unam.mx, romero@fisica.unam.mx}

\date{\today}

\begin{abstract}
Predictions from Hartree-Fock (HF), Popov (P), Yukalov-Yukalova (YY) and $t$-matrix approximations regarding the thermodynamics from the normal to the BEC phase in weakly interacting Bose gases are considered. By analyzing the dependence of the chemical potential $\mu$ on temperature $T$ and particle density $\rho$ we show that none of them predicts a second-order phase transition as required by symmetry-breaking general considerations. In this work we find that  the isothermal compressibility $\kappa_{T}$ predicted by these theories does not diverge at criticality as expected in a true second-order phase transition. Moreover the isotherms $\mu=\mu(\rho,T)$ typically exhibit a non-singled valued behavior in the vicinity of the BEC transition, a feature forbidden by general thermodynamic principles. This behavior can be avoided if a first order phase transition  is appealed. The facts described above show that although these mean field approximations give correct results near zero temperature they are endowed with thermodynamic anomalies in the vicinity of the BEC transition. We address the implications of these results in the interpretation of current experiments with ultracold trapped alkali gases.
\end{abstract}


\maketitle

\section{Introduction}
In the recent past it has been widely discussed that Bose-Einstein Condensation (BEC) in interacting Bose systems, i.e, the passage from the normal to the condensed phase where all particles occupy a single-particle state, shows an spontaneous $U(1)$ gauge symmetry breaking with the condensate fraction $N_{0}/N$ playing the role of the order parameter\cite{Lieb,Shi,Andersen,Yukalov}. The overwhelming task faced when trying to solve the full interacting quantum gas has motivated the search of physical approximations for the dilute and weakly interacting gas \cite{LandauII,Fetter}. This search has been also encouraged by the experimental realization of BEC in alkali gases\cite{Cornell1, Ketterle1} since BEC occurs in the regime of $s$ wave scattering where the interaction potential $U (\vec{r}_{1}-\vec{r}_{2})$ may be approximated by a contact potential $U (r_{1}-r_{2})=U_{0} \delta^{3}(r_{1}-r_{2})$\cite{Dalfovo}. The introduction of this simplification has lead to a formulation valid at low densities and temperatures near $T=0$. This approximation formally known as the theory of weakly interacting Bose gases has become the standard tool for analyzing the BEC transitions for gases confined either in a box of volume $V$ or trapped through external potentials in optical setups.\\

In spite of the great success achieved, the theory of weakly interacting Bose gas still awaits for a formal and complete analytical solution. In order to extract some partial results different approximations have been developed in an upper `layer'  build up on top of the theory of weakly interacting gases. This `layer' contains additional approximations to the already approximated theory of weakly interacting gases \cite{Shi,Goldman,Pethick,YY,Stoof,HYL,Reatto}.  In this work we focus our attention to four of the approximations most widely used to describe the interacting Bose gas in the dilute approximation. These are Hartree-Fock (HF), Popov (P), Yukalov-Yukalova (YY) and $t$-matrix approximations. As we show in this work none of these approximations exhibit BEC as a second-order phase transition. The fact that HF neither exhibits a second-order phase transition nor displays the correct energy spectrum has been known for a while \cite{Goldman,Pethick}. Here, we show that Popov (P), $t$-matrix approach\cite{Shi,Stoof} and the recently introduced Yukalov-Yukalova (YY) \cite{YY} approximations also fail to predict a second-order phase transition. In addition we show that all of these theories display an unstable region implying that the BEC transition is at best, a first-order phase transition.\\

To achieve this goal we proceed as follows. We start by calculating the equation of state $\mu = \mu(T,\rho)$,  that is,  the dependence of the chemical potential $\mu$ on temperature $T$ and particle density $\rho = N/V$, with $N$ the number of particles and $V$ the volume of the system in each of the approximations considered. Equilibrium thermodynamics ensures that the isothermal compressibility $\kappa_T$ can be obtained from the equation of state  $\mu = \mu(T,\rho)$ through the relationship $\kappa_T^{-1}= \rho^2 (\partial \mu / \partial \rho)_T$ from which information
 regarding the order of the transition predicted by these theories can be inferred. Since the isothermal compressibility $\kappa_T$ is indeed a thermodynamic
  manifestation of the fluctuations in the particle density, isotherms $\mu = \mu(T,\rho)$ contain fundamental information on the nature
  of the phase transition involved \cite{Callen}. \\

From a more general point of view the equation of state $\mu = \mu(\rho,T)$ of an interacting quantum gas contained in a rigid vessel of volume $V$ has a fundamental relevance on itself. Such a quantity enables the prediction of the density profile $n(\vec r)$ of the interacting gas confined by an inhomogenous trap potential $V_{ext}(\vec r)$ which is the main measurable property in the current experiments with ultracold alkali gases\cite{Cornell1,Ketterle1,Dalfovo,BrazilBEC}. The connection between the thermodynamics of the gas in a box of volume $V$ and the thermodynamics of the gas confined within an external potential $V_{ext}(\vec r)$ can be achieved through the Local Density Approximation (LDA). In such procedure one obtains $\rho = \rho(\mu,T)$ by inversion of the equation of state, and then replacing $\mu$ by $\mu - V_{ext}(\vec r)$. This gives the density profile $n(\vec r) = \rho(\mu - V_{ext}(\vec r),T)$.  This procedure has been shown to be exact in the appropriate thermodynamic limit of a gas confined by a trap $V_{ext}(\vec r)$\cite{Garrod,Marchioro,VRomeroC}. As we shall carefully discuss, if the trapped gas is in the BEC phase the density profile bears the information of the isotherm $\rho(\mu,T)$ above, below and at the transition. Thus, knowledge of $\mu = \mu(\rho,T)$ also yields the equation of state for the trapped gases. Inaccurate calculations of the equation of state in the homogenous case will be inherited to the inhomogeneous ones. One of the purposes of this article is to suggest high-resolution measurements of the density profiles in the current experiments. These would not only settle the issue of the validity or not of the mean-field calculations but would also pave the way to improve theoretical descriptions and thus better understanding of the BEC phase.\\

The paper is thus organized as follows. In Section II a general discussion of the thermodynamics of phase transitions and the weakly interacting gas is presented. Section III is devoted to a brief presentation of the approximations considered in this work and the results for the equation of state $\mu = \mu(T,\rho)$ corresponding to each of them.  Section IV discusses the implications of these results within the general theory of the weakly interacting gas and in the interpretation of current experiments in ultracold alkali gases.\\

\section{General thermodynamic considerations and the weakly interacting Bose gas}

For our analysis, we consider the equation of state $\mu = \mu(T,\rho)$,  that is,  the dependence of the chemical potential $\mu$ on temperature $T$ and particle density $\rho = N/V$, with $N$ the number of particles and $V$ the volume of the sample.  The extensivity property of Helmholtz free energy $F = F(N,V,T)$ allows us to either write $F = V f(\rho,T)$ or $F = N \tilde f(v,T)$ with $v = V/N = \rho^{-1}$ such that $f = \rho \tilde f$, and both forms carry the same physical information. One can  either calculate the chemical potential as
\begin{equation}
\mu(T,\rho) = \left( \frac{\partial f}{\partial \rho} \right)_T
\end{equation}
or the pressure
\begin{equation}
p(T,v) = -\left( \frac{\partial \tilde f}{\partial v} \right)_T .
\end{equation}
 Therefore, $\mu = \mu(T,\rho)$ and $p = p(v,T)$ also carry the same information and the phase diagram may be inferred from any of these forms. The laws of thermodynamics require that both of those expression are single valued, that is, for any given values of $\rho$ and $T$, or $v$ and $T$, there must only exist a single value of $\mu$ or $p$ respectively. In addition, these formulae carry out crucial information on the order of the phase transition involved.
 The stability of a particular thermodynamic state is ensured if the isothermal compressibility $\kappa_T$ is positive, where $\kappa_T$ may be either calculated as
\begin{equation}
\kappa_T^{-1} = \rho^2 \left(\frac{\partial \mu}{\partial \rho} \right)_T
\end{equation}
or
\begin{equation}
\kappa_T^{-1} = - v \left(\frac{\partial p}{\partial v} \right)_T .
\end{equation}
 These equations imply that $(\partial \mu / \partial \rho)_T  > 0$ and $(\partial p / \partial v)_T  < 0$ for any thermodynamic equilibrium state. An exception is a critical point where the latter derivatives become zero.\\

 The emergence of a negative region for the isothermal compressibility signals the onset of a first-order phase transition where two phases with different densities,  entropies and energies  coexist at the same temperature, pressure and chemical potential\cite{Callen}. Let us consider here that such states have densities $\rho_1$ and $\rho_2$. The equal-areas Maxwell construction enables us to calculate the values of the thermodynamic properties of interest for both coexisting states. In order to see this, it is enough to consider the Gibbs-Duhem relationship:
\begin{equation}
- S dT + V dp - N d\mu = 0,\label{GD}
\end{equation}
where $S$ is the entropy. By assuming the system is analyzed along an isotherm, we can set $dT \equiv 0$ and look for the coexisting values of the density or the volume. Then, we can either impose equal chemical potential $\mu = \mu_{coex}$ at both phases yielding
\begin{eqnarray}
\int_1^2 \> v dp & = & \int_1^2 \> d \mu = 0 \nonumber \\
&= & p_{coex} (v_2 - v_1) - \int_{v_1}^{v_2} p dv = 0, \label{MEA1}
\end{eqnarray}
where $p_{coex}$, $v_1$ and $v_2$ are the volume values at coexistency, with $v_1 = \rho_1^{-1}$ and $v_2 = \rho_2^{-1}$. The value of the chemical potential is $\mu_{coex} = \mu(T,\rho_1) = \mu(T,\rho_2)$. Equation (\ref{MEA1}) is the usual Maxwell equal-area construction in the $p - v$ diagram. Alternatively, if one imposes equal pressures at coexistence, namely $p = p_{coex}$, one finds from Eq.(\ref{GD})
\begin{eqnarray}
\int_1^2 \> \rho d\mu & = & \int_1^2 \> d p = 0 \nonumber \\
&= & \mu_{coex} (\rho_2 - \rho_1) - \int_{\rho_1}^{\rho_2} \mu d\rho = 0, \label{MEA2}
\end{eqnarray}
with $\mu_{coex}$, $\rho_1$ and $\rho_2$ the corresponding values at coexistency. Equation (\ref{MEA2}) is also a Maxwell equal-areas construction but in the $\mu - \rho$ diagram.\\

The onset of a second-order phase transition is featured by the divergence of the isothermal compressibility $\kappa_T$ and the heat capacity $C_v$ characterized by universal exponents\cite{Callen,Landau,Ma}. In this work we focus our attention to the isothermal compressibility  $\kappa_T$ whose divergence can be expressed by the vanishing derivative $(\partial \mu / \partial \rho)_T = 0$ or $(\partial p / \partial v)_T = 0$, at the transition. This condition demands that the isotherm $\mu = \mu(T,\rho)$ must become ``flat" at the critical density $\rho = \rho_c$. We shall explicitly show that none of the analyzed approximations for the weakly interacting Bose gas show this strong requirement. It is worth recalling that  an ideal  Bose gas confined in a box of volume $V$ does show a diverging compressibility at the critical temperature $T_c$\cite{Landau,Huang}.\\

The theory of the weakly interacting Bose gas assumes that the atoms have no structure since they typically are in the same hyperfine state\cite{Dalfovo}. If a pairwise interatomic potential  $U = U(|\vec r_i - \vec r_j|)$ is assumed the Hamiltonian in second quantization can be written as
\begin{equation}\label{Hsq}
\hat{H}= \sum_{k} \epsilon_{k}^{0} a^{\dagger}_{k} a_{k}+\frac{1}{2V} \sum_{k,k^\prime,q} \tilde U(q) a^{\dagger}_{k+q} a^{\dagger}_{k^\prime - q} a_{k} a_{k^\prime},
\end{equation}
where $k$ is the three dimensional wavevector quantized in a box of volume $V$, $a_k^\dagger$ and $a_k$ are creation and annihilation operators of particles with momentum $\hbar k$,  $\epsilon_k^0 = \hbar^2 k^2/2m$ is the one-particle kinetic energy, and $\tilde U(q)$ is the Fourier transform of the interparticle potential $U(r)$. The main assumption for the description of an interacting gas at low temperatures is the  contact potential approximation, namely, $\tilde U (q )\approx U_{0}$ for all $q$, with $U_{0}= 4 \pi \hbar^{2} a / m$ representing the strength of the interaction and $a$ the $s$-wave scattering length which in this work we shall assume as positive. In the $t$-matrix approximation also considered here, corrections to this simple potential are also included\cite{Shi}. An additional ansatz completes the framework. A Bose gas is considered weakly interacting if $\rho a^3 \ll 1$. All the approximations considered in this work fgall satisfy these general requirements. In the following we succinctly present the differences between each of them that yield to different equations of state $\mu = \mu(\rho,T)$. We refer to the reader to the original sources cited along this work for further details.\\

For our calculations we use units $\hbar = m = a = 1$. We shall analyze three isoterms in HF, P and YY approximations, $k_BT = $ 0.01,  0.1 and 1.1 in dimensionless units, while only $k_BT = $ 0.1 in TM approach. Before proceeding to the calculations it is worth to state some words on the physical regime associated to the isotherms chosen. Typical experiments on $^{23}$Na ultracold gases\cite{BrazilBEC} show that BEC transition temperature is nearly $T_c \approx 100$ nK with a scattering length $a \approx 55 a_0$, where $a_0$ is Bohr radius. This yields in dimensionless units $k_BT \approx 10^{-4}$, thus indicating that the chosen temperatures are a bit our of range, but this is not necessarily so. As we shall discuss below, while in HF and P the behavior is qualitatively the same for all temperatures, YY does show three different qualitative behaviors corresponding approximately to those temperatures. Moreover, recent advances in the experimental achievement of  ultracold atomic gases have shown that the scattering length $a$ can be tuned by external magnetic fields to larger values near a Feshbach resonance \cite{FeschKett,FeschCor,FeschHul,FeschIng}. Hence, if  an increase of tenfold in the scattering length is achieved,  then a temperature of $T \approx 100$ nK would give  $k_BT \approx $ 0.01 in dimensionless units which corresponds perfectly to the first isotherm considered in this work. The tunability of the scattering length $a$ together with with the fact that each time is  possible to achieve condensates with a larger number of particles suggest that in the near future BEC in ultracold gases will be achieved for higher temperatures.  Its is important to underline that the criterion for a weakly interacting gas is fulfilled here. The gas parameter associated to the isotherms are $\rho a^3 =$ 0.00017, 0.0052 and 0.19, respectively. Admittedly, the last value being at the border of what one should consider the gas as weakly interacting.

\section{BEC transition within four mean-field theories}

In this section we analyze four different approximations that describe the weakly interacting Bose gas. These schemes have provided accurate descriptions at and near zero temperature. Given this success their use have been extended to finite temperatures near criticality.  As suggested before, BEC can be attained either at fixed density lowering the temperature or at fixed temperature increasing the density. In this work, we shall make use of the latter approach and thus BEC transition will be accomplished in $\mu - \rho$ space at fixed temperature $T$ by varying the total density. In this case, the transition occurs at a critical density $\rho_c$, the gas being normal for $\rho < \rho_c$ while BEC sets in for $\rho \ge \rho_c$. \\

 \subsection{Hartree-Fock}

 The Hartree-Fock approximation is a self-consistent approach in which the state of the $N-$particle system is expressed in terms of effective one-particle states, yielding a gas of non-interacting excitations whose energy spectrum depends self-consistently on its density and the actual interparticle interaction\cite{Goldman,Pethick}. Then, one can construct the grand potential as $\Omega = \langle H \rangle - TS + \mu \langle N \rangle$, in terms of variational occupation numbers $f_q$, with
 \begin{equation}
  \langle H \rangle \approx \frac{N_{0}^{2} U_{0}}{2 V}+ \sum_{k \neq 0} \left( \epsilon_{k}^{0}+ 2 \rho U_{0}\right)f_{k} - \frac{U_{0}}{V}\sum_{k, q\ne 0}f_{k}f_{q} \label{EHF}
\end{equation}
and
\begin{equation}
  \langle N \rangle \approx N_0 + \sum_{k \ne 0} f_k
  \end{equation}
with $N_0$ the number of particles in the condensate with $k \equiv 0$. Further, one assumes that the entropy is given as that of an ideal Bose gas, but in terms of the effective one-particle occupation states $f_k$, this is 
$S = k \sum_k \left[ (1 + f_k) \ln(1 + f_k) - f_k \ln f_k \right]$.  At zero temperature, HF is considered an adequate approximation being equivalent to the Gross-Pitaevskii approach. However, at low but finite temperatures it predicts an excitation spectrum with a gap  $\epsilon_k = \epsilon_k^0 + 2 \rho U_0$ as seen from Eq.(\ref{EHF}) contrary to the expected gapless spectrum\cite{Bogoliubov}, that P and YY approximations do take into account. However, in the normal gas region, that is, for densities below and at the transition, HF is an acceptable approximation. \\

By looking for the values of $f_k$ that minimize $\Omega$ at constant temperature, one is lead to the thermodynamic equations in HF approximation. One finds that for  densities $\rho <  \rho_{c}$, the gas is in the normal phase with\cite{Goldman,Pethick} 
\begin{equation}\label{Eq1}
\rho  = \frac{1}{\lambda_{T}^{3}} g_{3/2} \left[ \beta \left( \mu-2 \rho U_{0}\right) \right] .
\end{equation}
On the other hand, for $\rho \ge \rho_{c}$ BEC sets in, and it is found that,
\begin{equation}\label{Eq2}
\rho= \rho_{0}+\frac{1}{\lambda_{T}^{3}}g_{3/2} \left( - \beta \rho_{0} U_{0} \right) \hspace{0.5cm}  \mbox{with} \hspace{0.5cm} \mu=\left( 2 \rho -\rho_{0}\right)
U_{0} ,
\end{equation}
where $\rho$ and $\rho_0$ are the total and condensate densities. In the above equations $\lambda_{T}= h/{\sqrt{2 \pi m k_{B} T}}$, and we used the Bose integral
\begin{equation}\label{Equation2.5}
g_{n} (\alpha) =\frac{1}{\Gamma (n)} \int_{0}^{\infty} \frac{x^{n-1}}{e^{x - \alpha}-1}dx ,
\end{equation}
with $\beta = 1/k_BT$. \\

The onset of BEC is assumed to occur at $\rho_0 = 0$ in Eq.(\ref{Eq2}), namely at $\mu_c = 2 \rho_c  U_0$. This defines a relation between $T$ and $\rho_c$, which is the same as in the ideal gas,
\begin{equation}
\rho_c = \frac{1}{\lambda_{T}^{3}} g_{3/2}(0) .\label{BEC-cond}
\end{equation}
Eqs. (\ref{Eq1}) and (\ref{Eq2}) yield $\mu = \mu(\rho,T)$ as a continuous function of $\rho$ at BEC. As we now show, this continuity does not guarantee the existence of a second-order phase transition. \\

\begin{figure}
  \begin{center} \scalebox{0.8}
   {\includegraphics[width=\columnwidth,keepaspectratio]{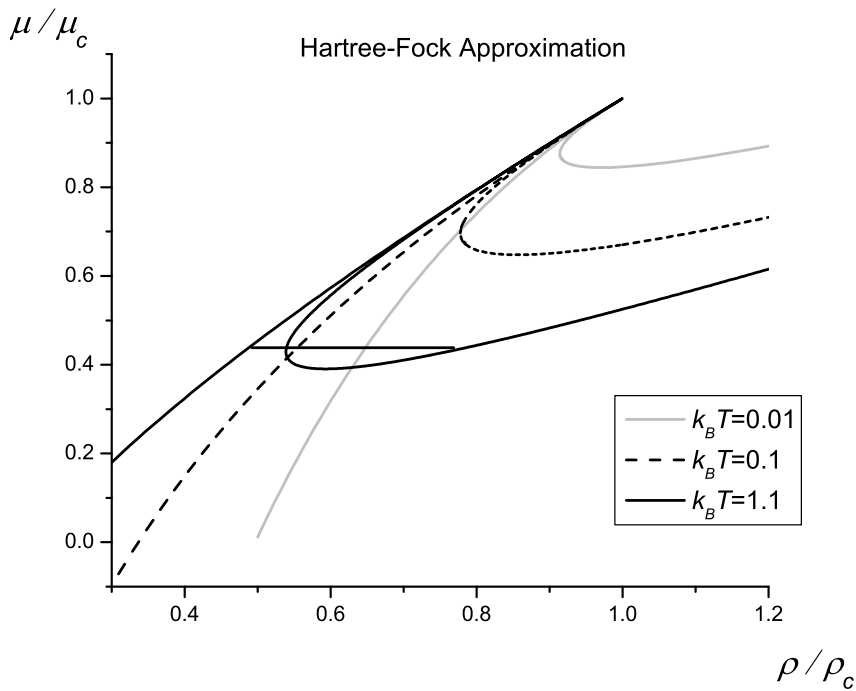}}
   \caption{Isotherms $\mu$ vs $\rho$ at $k_BT =$ 0.01, 0.1 and 1.1, from numerical solutions of HF equations (\ref{Eq1}) and (\ref{Eq2}). $\mu_{c}$ and $\rho_{c}$ are the values at BEC. The gaseousness parameter at criticality are $\rho_c a^3 =$ 0.00017, 0.0052 and 0.19, respectively. The isotherm $k_B T= 1.1$ shows Maxwell equal-areas construction for a first-order transition.}\label{uno}
  \end{center}
\end{figure}

Fig. \ref{uno} shows solutions to Eqs.(\ref{Eq1}) and (\ref{Eq2}) for the three isotherms, $k_BT = $ 0.01,  0.1 and 1.1. The behavior of $\mu$ vs $\rho$ for all temperatures is essentially the same. In the normal gas region, $\rho < \rho_c$, the derivative $(\partial \mu/\partial \rho)_T \to 2 U_0$ as $\rho \to \rho_c$. This straightforwardly predicts that the isothermal compressibility does not diverge at BEC. The behavior of  $\mu$ vs $\rho$ corresponding to the BEC  region, i.e. the solution to Eq.(\ref{Eq2}),  shows an anomalous thermodynamic behavior: this solution, in principle only valid for densities $\rho \ge \rho_c$, intrudes into the normal region yielding a multiple valued chemical potential as a function of density. Thermodynamics forbids this multiple-valuedness. However, as done originally in Refs.\cite{HYL} and \cite{Reatto}, such a behavior can be avoided by invoking a first-order phase transition, joining two phases with different densities by means of an equal-areas Maxwell construction, as shown in Fig. \ref{uno} for the isotherm $k_BT = 1.1$. As discussed above, equal-areas Maxwell construction ensures that the two phases with different densities have the same temperature, pressure and chemical potential, namely, that the phases coexist. This is the signature of a first order phase transition and one finds that the conditions for a second-order phase transition, as discussed before are never met. As described previously the fact of having the coexistence of two states with different densities implies two different entropies, namely a latent heat, and two values of the energy. In the same fashion, a multiple valued chemical potential gives rise to also an unacceptable multiple-valued pressure of the system.

\subsection{Popov}

The Popov approximation is considered correct at zero temperatures being essentially the same as Bogoliubov approximation\cite{Bogoliubov}.  This approximation, as shown below, yields a gapless excitation spectrum\cite{Shi,Andersen,Pethick}, linear in the excitation momentum as $k \to 0$, and that gives rise to the phenomenon of superfluidity, just as in $^4$He\cite{Fetter}.
Bragg spectroscopy in an ultracold $^{87}$Rb gas\cite{BEC-spectrum} has shown that the excitation spectrum is clearly given by the Bogoliubov expression, thus providing validity to this approximation at very low temperatures and, incidentally, confirming the superfluid nature of these gases at the BEC state.\\

The number-conserving Hamiltonian of the system at this level of approximation is given by\cite{Pethick}
\begin{eqnarray}
\hat H &\approx&  \frac{N_0^2}{2V} + \sum_{k}  \left( \epsilon_k^0 + \frac{2 N U_0 }{V} \right) \>  a_{k}^\dagger a_k \> - \frac{U_{0}}{V}\sum_{k, q\ne 0}f_{k}f_{q} +  \nonumber \\
&& \frac{N_0 U_0}{2 V} \sum_{k \ne 0} \left( a_{k}^\dagger a_{-k}^\dagger + a_k a_{-k} \right) .\label{BH}
\end{eqnarray}
It looks very similar to the Hartree-Fock version except for the last term which represents the annihilation of two particles into the condensate and the creation of two particles from the condensate. For a finite temperature calculation of the thermodynamics once again one deals with the problem in the grand canonical ensemble. After performing the usual Bogoliubov transformation\cite{Pethick}, the Popov procedure follows a similar line as HF and the system emerges as a kind of ideal gas with  elementary excitations whose excitation spectrum is,
\begin{equation}\label{excP}
\epsilon_{k}= \sqrt{ \left[ \frac{\hbar^2 k^{2}}{2m}+2 \rho U_{0} - \mu \right]^{2}-\rho_{0}^{2}U_{0}^{2}}.
\end{equation}

\begin{figure}
  \begin{center} \scalebox{0.8}
   {\includegraphics[width=\columnwidth,keepaspectratio]{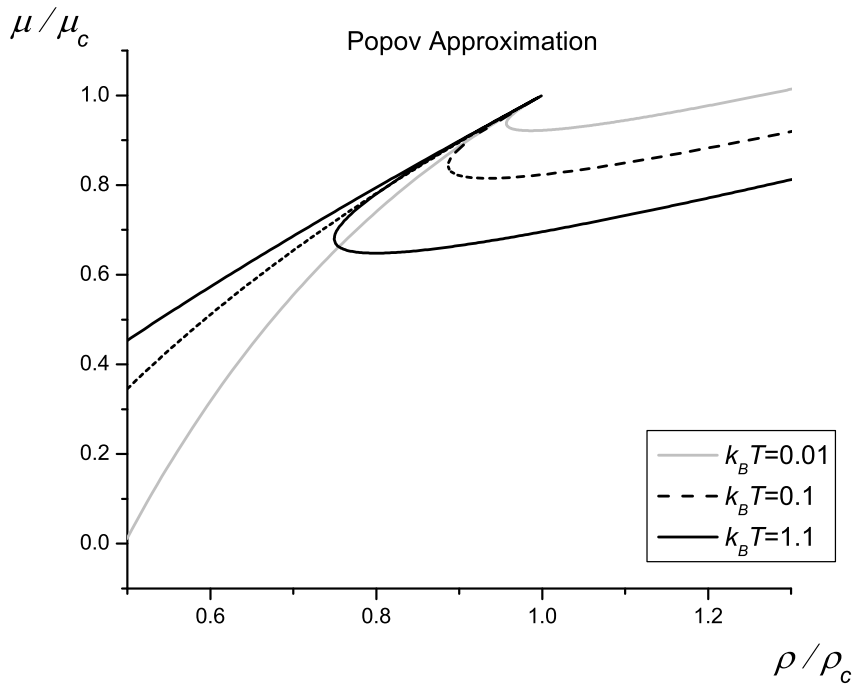}}
   \caption{Isotherms $\mu$ vs $\rho$ at $k_BT =$ 0.01, 0.1 and 1.1, from numerical solution of Popov equations (\ref{Eq1}) and (\ref{Eq3}). $\mu_{c}$ and $\rho_{c}$ are the values at BEC.  The gaseousness parameter at criticality are $\rho_c a^3 =$ 0.00017, 0.0052 and 0.19, respectively. }\label{cuatro}
  \end{center}
\end{figure}

At a fixed temperature $T$ and for densities below the critical one $\rho_c$, the equation in P approximation for the density $\rho$ as a function of  temperature $T$ and chemical potential $\mu$, turns out to be the same as in HF, Eq.(\ref{Eq1}), but for densities $\rho \ge \rho_c$ it is found that the density of the fluid is
\begin{equation}\label{Eq3}
\rho = \rho_{0} + \frac{2}{ \pi^{1/2} \lambda_T^3} \int_{0}^{\infty}\frac{   \left(x + \beta \rho_{0} U_{0} \right) }{\sqrt{x^2 + 2  \beta \rho_{0} U_{0} x }} \frac{x^{1/2} dx}{e^{\sqrt{x^2 + 2  \beta \rho_{0} U_{0} x }}-1}
\end{equation}
while the chemical potential is again given by $\mu = ( 2 \rho- \rho_{0}) U_{0} $ as in HF. The transition occurs when the condensate density vanishes, $\rho_0 = 0$, and the critical density $\rho_c$  is given by the condition (\ref{BEC-cond}). The solution $\mu(\rho,T)$ for $\rho \ge \rho_c$ becomes continuous at $\rho_c$ with the solution to Eq.(\ref{Eq1}). Fig. \ref{cuatro} shows three isotherms $\mu$ vs $\rho$ obtained fo the Popov approximation using Eqs.(\ref{Eq1}) and (\ref{Eq3}) for the same values used before. It is  found that although the multiple-valued region is smaller in Popov description than in Hartree-Fock the thermodynamic behavior is qualitatively similar. An equal-areas Maxwell construction would yield to a first-order phase transition as well.

\subsection{Yukalov-Yukalova}

This recent scheme is based on the inclusion of anomalous averages $\sigma_1 = \langle \hat \psi_1(\vec r) \hat \psi_1(\vec r) \rangle$ into the Hamiltonian. According to its authors the anomalous averages should not be neglected since their contribution is of the same order as that of the condensate density \cite{YY}. The operator  $\hat \psi_1(\vec r)$ is given in terms of the creation operators $a_k$ as
\begin{equation}
\hat \psi_1(\vec r) = \sum_{k \ne 0} \phi_k(\vec r) a_k .
\end{equation}
 It is discussed in Ref. \cite{YY} that these anomalous are not properly considered in Popov approximation and therefore their inclusion provides with a better description for the BEC transition. In order to obtain the thermodynamics of the system YY approximation considers explicitly the symmetry breaking of the state by the appearance of the condensate and implements a so-called representative ensemble calculation where one Legendre multiplier is associated to the condensate particles $\rho_0$ and a different one to the non-condensate fraction $\rho - \rho_0$. This takes into account correlations of uncondensed particles and allows for a clear distinction of the superfluid and condensate fractions at finite temperatures. At zero temperature, this approximation also agrees with that of Bogoliubov\cite{Bogoliubov} as expected. The details of the calculation can be consulted in the original works of Yukalov and Yukalova\cite{YY}.\\

 Once more, nevertheless, for a given temperature $T$ and in the normal region $\rho < \rho_c$,  the relevant equations are those of HF, Eq.(\ref{Eq1}). For densities above criticality, $\rho \ge \rho_c$, a new set of equations are given for the the value of the density $\rho$ and the anomalous average $\sigma_1$ namely,
\begin{eqnarray}\label{YY1}
\rho &=&\rho_{0}+\frac{1}{3 \pi^{2}} \left(\frac{m c}{\hbar}\right)^3 \left[  1+ \frac{3}{2 \sqrt{2}} \int_{0}^{\infty}
\left( \sqrt{1+x^{2}}-1\right)^{1/2}  \nonumber \right.\\
&&\times  \left.  \left[ \coth{\left(\frac{mc^2 x}{2 k_BT}  \right)}-1 \right]dx \right]
\end{eqnarray}

\begin{eqnarray}\label{YY2}
\sigma_{1}&=& \left(\frac{m c}{\hbar}\right)^3 \left[ \left(\frac{ \rho_{0} U_{0}}{\pi^4 mc^2}\right)^{1/2}-
\right.  \frac{1}{2 \pi^{2}\sqrt{2}} \times \nonumber \\
&& \left. \int_{0}^{\infty}
\frac{\left( \sqrt{1+x^{2}}-1\right)^{1/2}}{\sqrt{1+x^{2}}} \left[ \coth{\left(\frac{mc^2 x}{2k_BT}  \right)}-1 \right]dx \right],
\end{eqnarray}
where the speed of the elementary excitations $c$ is given in terms of the condensate density $\rho_0$ and $\sigma_1$ by $mc^{2} = \left( \rho_{0}+\sigma_{1} \right)U_{0}$. In this approximation, the chemical potential is given by\cite{YY}
\begin{equation}
\rho \mu= \rho_{0} \left(2 \rho - \rho_{0}+ \sigma_{1}\right)U_{0} +  (\rho-\rho_{0}) \left( 2 \rho - \rho_{0} - \sigma_{1} \right)U_{0} . \label{YY3}
\end{equation}

\begin{figure}
  \begin{center} \scalebox{0.8}
   {\includegraphics[width=\columnwidth,keepaspectratio]{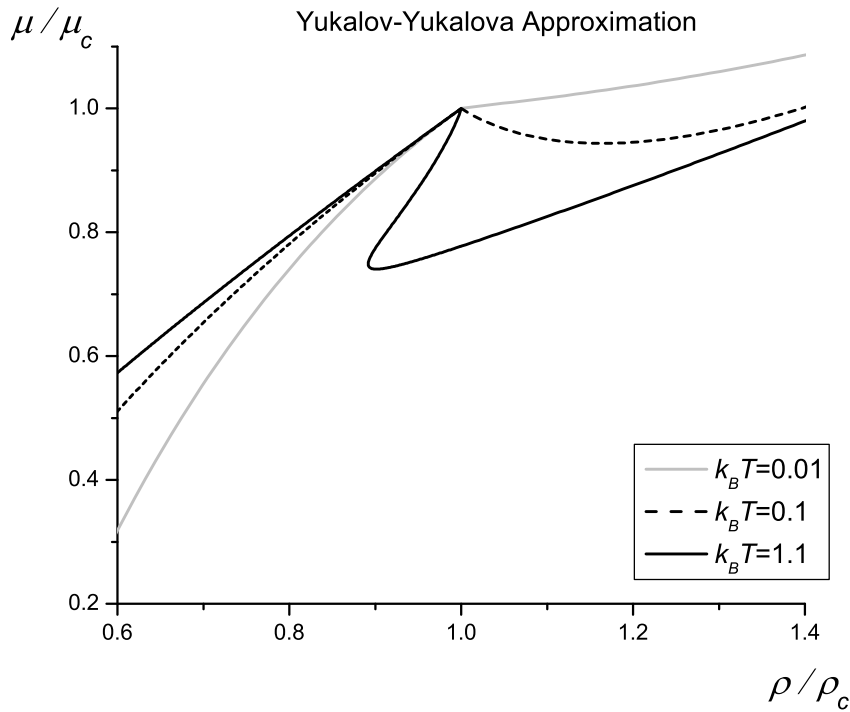}}
   \caption{Isotherms $\mu$ vs $\rho$ at $k_BT =$ 0.01, 0.1 and 1.1, from numerical solution of YY equations (\ref{Eq1}) and (\ref{YY1})- (\ref{YY3}). $\mu_{c}$ and $\rho_{c}$ are the values at BEC. The gaseousness parameter at criticality are $\rho_c a^3 =$ 0.00017, 0.0052 and 0.19, respectively.}\label{seis}
  \end{center}
\end{figure}

In this case BEC sets in when both $\sigma_1$ and $\rho_0$ vanish. This yields $\mu_c = 2 \rho_c U_0$ at the transition with $\rho_c$ determined by the condition (\ref{BEC-cond}). As in HF and P, the solution $\mu(\rho,T)$ to (\ref{YY1})-(\ref{YY3}) is continuous with the solution of (\ref{Eq1}) at BEC.
This approximation is much richer than the two previous ones. The behavior of the isotherms $\mu$ vs $\rho$ show three different regimes depending on the temperature, see Fig. \ref{seis}.  At very low temperatures, i.e. $k_BT = 0.01$, there are no multiple valued solutions, neither an unstable region. Thus, there is no need for appealing to a first-order phase transition. However, despite the fact that the transition appears continuous, it cannot be considered as a bona-fide second order one since neither side shows a divergent compressibility at BEC;  rather, the compressibility is discontinuous. Nevertheless, this case of YY is the closest to a second-order phase transition. At intermediate temperatures, i.e. $k_BT = 0.1$, although no multiple valued solutions exist, an unstable region appears where $(\partial \mu/\partial \rho)_T < 0$, and a first-order phase transition must therefore be adscribed. At large temperatures, $k_BT = 1.1$, the multiple-value issues of HF and P pervade YY as well.

\subsection{Many-Body $t$-matrix}

This approach was originally carried out in Ref.\cite{Stoof} but here we follow the analysis given in Ref. \cite{Shi}. This approximation builds on the Popov theory and takes into account higher corrections to the scattering $t$-matrix including many-body effects, thus going beyond the contact interaction potential $\tilde U(k) \approx U_0$.\\

This case does not have HF as the solution in the normal phase. Instead, for $\rho < \rho_c$ and fixed $T$, the density equation for the normal gas is,
\begin{equation}
\rho = \frac{1}{\lambda_T^3} \> g_{3/2}(\beta \Delta) ,\label{SG1}
\end{equation}
while for $\rho \ge \rho_c$,
\begin{equation}
\rho = \rho_0 + \frac{1}{\pi^{1/2} \lambda_T^3} \int_0^\infty \left[ \frac{x + \beta \Delta}{E_x} \coth \frac{E_x}{2} - 1 \right] x^{1/2} \> dx . \label{SG4}
\end{equation}
In the above equations, $\Delta$ is determined by an additional quantity,
\begin{equation}
\alpha =\frac{1}{\pi^{1/2} k_BT \lambda_T^3} \int_0^\infty \left[ \frac{1}{E_x} \coth \frac{E_x}{2} - \frac{1}{x} \right] x^{1/2} \> dx , \label{SG2}
\end{equation}
where, for $\rho < \rho_c$, $E_x = x - \beta \Delta$, and for $\rho \ge \rho_c$, $E_x = \sqrt{x^2 + 2 \beta \Delta x}$. These equations must be solved self-consistently with the corresponding equation for the chemical potential,
\begin{equation}
\mu = \Delta +  \frac{2 \rho U_0}{(1 + \alpha U_0)} . \label{TM3}
\end{equation}
 BEC occurs when $\Delta \to 0$ and $\alpha \to \infty$ from both sides.
This yields $\mu = 0$ at the transition, as in the ideal gas, and the transition density is again given by condition (\ref{BEC-cond}).\\

\begin{figure}
  \begin{center} \scalebox{0.8}
   {\includegraphics[width=\columnwidth,keepaspectratio]{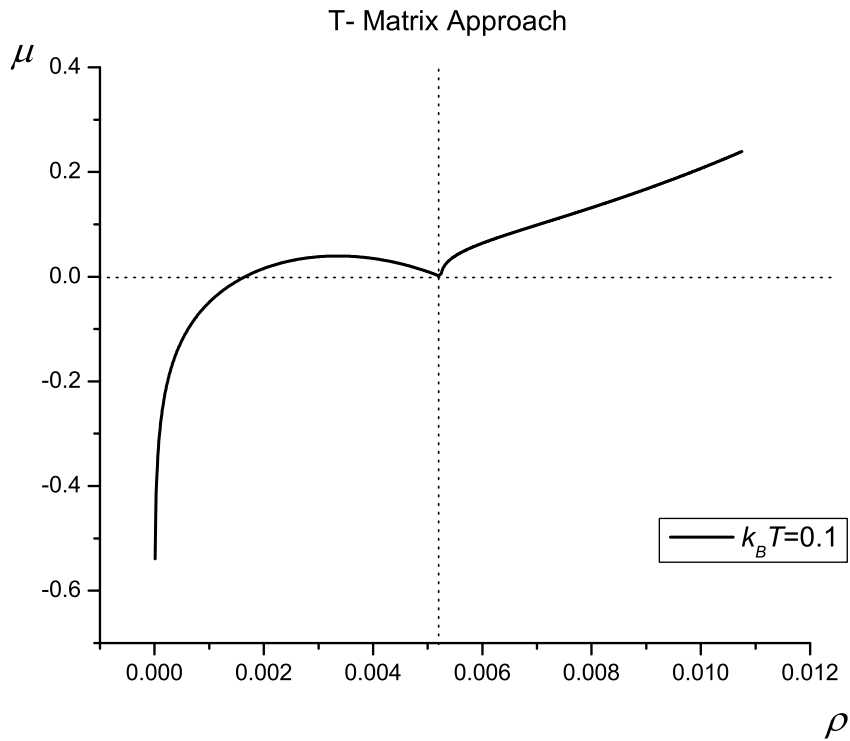}}
   \caption{Isotherm $\mu$ vs $\rho$ at $k_BT = 0.1$, from numerical solution of $t$ matrix equations (\ref{SG1})-(\ref{TM3}). At BEC this approximation predicts a full incompressibility of the gas. This property is completely unexpected and it is contrary to the standard results in weakly interacting systems.  }\label{ocho}
  \end{center}
\end{figure}

Figure \ref{ocho} shows the isotherm $k_BT = 0.1$, with $\rho_c a^3 = 0.0052$. Given the fact that this theory goes beyond mean-field by considering the renormalization of the coupling parameter, it is somewhat surprising that its predictions, in a way, fare less satisfactory than HP and P. It is seen that in the normal region $\rho < \rho_c$ the solution is unstable near BEC, i.e.  $(\partial \mu/\partial \rho)_T < 0$. In the BEC side,  $\rho \ge \rho_c$ the situation is more worrisome, since it appears that $(\partial \mu/\partial \rho)_T$ becomes extremely large, probably diverging at the transition. This indicates that the compressibility becomes nearly zero, that is, the gas becomes {\it incompressible} at BEC. This is completely opposite to a critical behavior where large density fluctuations are due to the large compressibility of the gas\cite{Ma}.

\section{Final remarks and Perspectives}

In this work we considered four of the most used mean field theories to describe the onset of the BEC phase in weakly interacting gases. We calculated the equation of state $\mu=\mu(\rho,T)$ corresponding to each of these theories at three different temperature regimes (with exception of the $t$-matrix theory which has calculated only at one temperature) satisfying the requirements of dilute gas in the contact potential approximation. The central conclusion is that none of the descriptions considered display a true second-order phase transition at BEC as expected by general symmetry breaking considerations. In addition and excepting the Yukalov-Yukalova approximation at $k_BT=0.01$ and 0.1 all the approximations considered exhibit an unphysical behavior in the vicinity of BEC. This anomalous property restricts severely its applicability at finite temperatures near the transition. \\

The interest in the equation of state equation of state $\mu = \mu(\rho,T)$ goes beyond the order of the phase transition involved. As discussed in the Introduction it provides a fundamental link to the determination of the local density profile $n(\vec r)$ in gases confined by inhomogeneous magnetic or optical traps which is one of the main measurable quantities in the current experiments with ultracold alkali gases\cite{Cornell1,Ketterle1,Bagnato1}.  It is worthwhile to point out that a {\it single} profile at  a  temperature below BEC in the trapped gas bears information of the {\it homogeneous} one  for densities $\rho$ below, at and above the critical one $\rho_c$.\\

\begin{figure}
  \begin{center} \scalebox{0.8}
   {\includegraphics[width=\columnwidth,keepaspectratio]{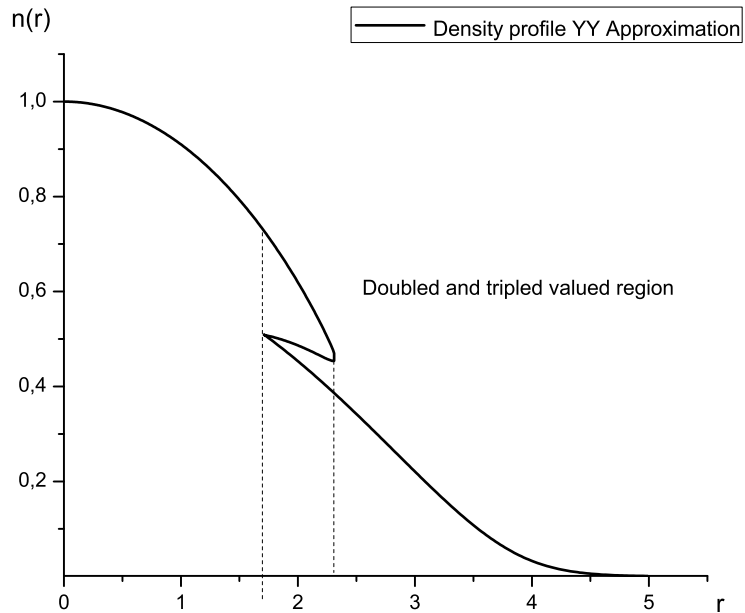}}
   \caption{Density profile $\rho(r)$ of a harmonically trapped Bose gas in the YY approximation at temperature $k_BT = 1.1$ obtained through the implementation of Local Density Approximation. }
\label{nueve}
  \end{center}
\end{figure}

Using the Local Density Approximation (LDA) we have calculated density profiles for an isotropic 3D harmonic trap, $V_{ext}(\vec r) = \frac{1}{2} m \omega^{2} r^{2}$ using the isotherms $\mu = \mu(\rho,T)$ obtained from the analyzed theories. In this case the thermodynamic limit corresponds to $N \to \infty$, $\omega \to 0$ but $N \omega^3 =$ constant. It is found that the predicted density profiles for HP and P are multiple valued for any temperature, while for YY this occurs only for relatively high temperatures. As an illustration of these multiple valuedness we show in Fig. \ref{nueve} the YY profile obtained for $k_BT = 1.1$ by applying LDA to the corresponding case in Fig. \ref{seis}. We recall that LDA, within the HF and P approximations, has been widely used to calculate thermodynamic properties of trapped gases, see e.g. Refs.\cite{Dalfovo,Goldman,Giorgini,Minguzzi,VRomeroA}, and the issue of multiple-valuedness has not been brought up. This is perhaps due to the fact that, when performing numerical calculations, the multiple valued region being very small can be ``short-circuited", inadvertently or not. In that case, the resulting profile is thus continuous and, apparently, the ensuing thermodynamic properties of the trapped gas are quite insensitive to this correction. We expect that the results provided in this work will bring again the attention to this matter. An accurate determination of density profiles would help to settle this issue and certainly it would provide solid guidelines to a better formulation of the theory of the weakly interacting gas.\\

Although the mean-field theories here described do not predict a second-order phase transition, their behavior near $T = 0$ remains correct and valuable. With the exception of HF, the others correctly incorporate the gapless elementary excitation spectrum, originally predicted by Bogoliubov\cite{Bogoliubov}. The results presented here suggest strongly that at temperatures near the transition additional properties of the interatomic potential are needed to fully capture the thermodynamics of the system. Nevertheless it is important to stress that exact calculations of the thermodynamics of quantum systems based on complete solutions of the Hamiltonian (\ref{Hsq}) are beyond present capabilities. We believe that in the construction of a new approximation or an improvement of those, the calculation of the equation of state $\mu = \mu(\rho,T)$ should be a useful tool to determine the predicted order of the transition, and therefore, of the correct thermodynamic behavior of the ultracold gas in the BEC state.

\ack{
We acknowledge support from grants PAPIIT-UNAM IN114308 and IN116110. L. Olivares-Quiroz gratefully acknowledges support from CONACyT and Universidad Autonoma de la Ciudad de Mexico}

\end{document}